\documentclass[conference]{IEEEtran}
\IEEEoverridecommandlockouts
\usepackage{cite}
\usepackage{amsmath,amssymb,amsfonts}
\usepackage{algorithmic}
\usepackage{graphicx}
\usepackage{textcomp}
\usepackage{xcolor}
\usepackage{hyperref}

\usepackage{cite}
\usepackage{amsmath,amssymb,amsfonts}
\usepackage[normalem]{ulem}
\usepackage{multirow}

\def\BibTeX{{\rm B\kern-.05em{\sc i\kern-.025em b}\kern-.08em
    T\kern-.1667em\lower.7ex\hbox{E}\kern-.125emX}}
\begin{document}

\title{An Analytical Estimation of Spiking Neural Networks Energy Efficiency\\
\thanks{This research is funded by the ANR project DeepSee, Université Côte d’Azur, CNRS and Région Sud Provence-Alpes-Côte d’Azur.}
}


\author{\IEEEauthorblockN{Edgar Lemaire}
\IEEEauthorblockA{\textit{LEAT, Univ. C\^ote d'Azur, CNRS} \\
edgar.lemaire@univ-cotedazur.fr}
\and
\IEEEauthorblockN{ Lo\"ic Cordone}
\IEEEauthorblockA{\textit{Renault Software Factory} \\
\textit{LEAT, Univ. C\^ote d'Azur, CNRS}\\
loic.cordone@univ-cotedazur.fr}
\and
\IEEEauthorblockN{Andrea Castagnetti}
\IEEEauthorblockA{\textit{LEAT, Univ. C\^ote d'Azur, CNRS} \\
andrea.castagnetti@univ-cotedazur.fr}
\and
\IEEEauthorblockN{Pierre-Emmanuel Novac}
\IEEEauthorblockA{\textit{LEAT, Univ. C\^ote d'Azur, CNRS} \\
pierre-emmanuel.novac@univ-cotedazur.fr}
\and
\IEEEauthorblockN{Jonathan Courtois}
\IEEEauthorblockA{\textit{LEAT, Univ. C\^ote d'Azur, CNRS} \\
jonathan.courtois@univ-cotedazur.fr}
\and
\IEEEauthorblockN{Beno\^it Miramond}
\IEEEauthorblockA{\textit{LEAT, Univ. C\^ote d'Azur, CNRS} \\
benoit.miramond@univ-cotedazur.fr}
}

\maketitle

\begin{abstract}
Spiking Neural Networks are a type of neural networks where neurons communicate using only spikes. They are often presented as a low-power alternative to classical neural networks, but few works have proven these claims to be true. In this work, we present a metric to estimate the energy consumption of SNNs independently of a specific hardware. We then apply this metric on SNNs processing three different data types (static, dynamic and event-based) representative of real-world applications. As a result, all of our SNNs are 6 to 8 times more efficient than their FNN counterparts.
\end{abstract}


\section{Introduction}


Neuromorphic computing has been studied for many years as a game changer to address low-power embedded AI, assuming that the inspiration from the brain will natively come with a reduction in energy consumption. Neuromorphic computing mainly focuses on the encoding and the processing of the information with spikes. If this property takes an obvious place in the biological functioning, it is far from obvious that it is the only one to explain the efficiency of the brain. It is therefore necessary to ask the question whether considering this characteristic in isolation brings a gain compared to the classical neural networks used in deep learning. This is the question that this paper seeks to answer by restricting the study to standard machine learning tasks on three different types of data: static, dynamic and event-based data.

There already exist comparisons between Spiking Neural Networks (SNNs) and Formal Neural Networks (FNNs, \textit{i.e.} non-spiking Artificial Neural Networks) in the literature. However, such comparisons are hardly generalizable since they focus on specific applications or hardware targets \cite{confortingml}, \cite{NxTF}. Moreover, the considered applications are often toy examples not representative of real-world AI tasks. Another approach consists in producing metrics in order to evaluate the relative energy consumption between the two coding domains, based on their respective synaptic operations and activity. 
We thus propose a novel metric for energy consumption estimation taking synaptic operations, memory accesses and element addressing into account. Moreover, our metric is mostly independent from low-level implementation 
or hardware target to ensure its generality.


The proposed metric is described and applied to three datasets representative of the aforementioned data types: CIFAR-10 for the static case, Google Speech Commands V2 for the dynamic case, and Prophesee NCARS for the event-based case. Moreover, those datasets are closer to real-world applications than usual benchmarks of the neuromorphic community. The metric is used in conjunction with accuracy measurements to provide an in-depth evaluation of those three application cases and their relevance for spiking acceleration. We use the advanced Surrogate Gradient Learning technique and Direct Encoding spike conversion method, since they offer the best trade-off between prediction accuracy and synaptic activity \cite{lemaire2022modelisation}.

Our code and trained SNN models are available upon request. 

\section{State of the art}

\subsubsection{Data Encoding}
\label{sec-SOTA-Encoding}

In order to process data in SNNs, it must be encoded into spikes. Rate, Time and Direct encoding are three methods used to convert conventional data towards spiking domain. Rate coding \cite{lemaire2022synaptic} is the most notorious, since it provides state-of-the-art accuracy on most AI tasks. However, it generates a lot of spikes over a large number of timesteps, drastically impacting computational and energy efficiency. Time coding \cite{kheradpisheh2020temporal} intends to cope with this issue by encoding information in latency rather than rate, thus generating much fewer spikes. Yet, the temporal sparsity of latency-coded spikes causes long processing times, resulting in an energy overhead. 

To cope with those limitations, we address a novel encoding scheme: Direct Encoding \cite{zimmer2019technical}. It should be noted that this term is a proposition of ours. In Direct Encoding, the first processing layer is made of hybrid neurons with analog inputs and spiking behaviour (IF, LIF...). The weights of this layer are learned during training, thus encoding can be tuned to reduce spiking activity in the network \cite{pellegrini2021low}. In the present work, we evaluate Direct Encoding on CIFAR-10 and GSC datasets. Additionally, we evaluate native spike encoding using event cameras \cite{amir2017low}. In this method, each pixel of the sensor generates a spike whenever it detects a brightness variation, thus encoding movement into spikes. With such sensor, the input spiking activity is very low, since only the information of interest (\textit{i.e.} moving objects) are returned by the camera. This property helps improving the computational and energy efficiency of SNNs. We evaluate native encoding using the Prophesee NCARS dataset.

\subsubsection{Training of SNN in the literature}
\label{sec-SOTA-Training}


Spiking neural networks cannot use the classical backpropagation training algorithm to learn their weights because its activations (spikes) are binary and thus non-differentiable. Encoding static data such as images using rate coding enables the conversion of an already trained FNN to a SNN. The most common way is to replace the ReLU neurons of the FNN by IF neurons \cite{ijcai2021-321}. However, the prediction accuracy obtained through conversion is systematically inferior to their FNN counterpart, while generating a lot of spikes over a large number of timesteps. 
Numerous works have studied how to train SNNs directly in spiking domain. The best results were obtained using backpropagation-based learning rules, such as the surrogate gradient \cite{neftci}.
To circumvent the non-differentiability of spikes in SNNs, the main idea of surrogate gradient learning is to use two distinct functions in the forward and backward passes: an Heaviside step function for the first, and a differentiable approximation of the Heaviside in the latter, such as a sigmoid function.
Using surrogate gradient learning requires a fixed number of timesteps. As the number of computations performed by SNNs increases with the number of timesteps, being able to fix it beforehand and to tune it during the training is vital to increase the computational efficiency of SNNs. 
\subsubsection{Comparisons based on measurements} 

In the literature, a few comparisons of SNNs and FNNs have been produced based on hardware measurements. Some papers show competitive results for SNNs: in \cite{SNN4FBnEB}, the authors highlighted the influence of the spike encoding method on the accuracy and computational efficiency of the SNN. They compared the spiking and formal networks through a Resnet-18 like architecture on
two classification datasets. They found that SNNs reached higher or equivalent accuracy and energy efficiency. In \cite{confortingml}, the authors showed that an SNN could reach twice the power and resource efficiency of an FNN, with an MLP on MNIST dataset targeting ASIC. However, those encouraging results are still very specific and thus hardly generalizable. In a more holistic approach, the authors of \cite{lemaire2022synaptic} performed a design space exploration (including encoding, training method, level of parallelism...) and showed that the advantage of the SNN depended on the considered case, making it difficult to draw general rules. In \cite{NxTF}, researchers showed that SNNs on dedicated hardware (Loihi) demonstrated better energy efficiency than equivalent FNNs on generic hardware (CPU and GPU) for small topologies, but observed the opposite using larger CNNs. Once more, the conclusions depended on the studied case and could not be generalized. Albeit encouraging, those results are not sufficient to draw general conclusions regarding the savings offered by event-based processing, since they depend on the selected application, network hyper-parameters and hardware targets. Therefore, another approach consists in comparing both coding domains through estimation metrics, taking a step back to produce more general conclusions.





\subsubsection{Comparisons based on metrics}
Most energy consumption metrics are based on the number of synaptic operations: accumulations (ACC) in the SNN and multiplication-accumulations (MAC) in the FNN. Those models have limitations: energy consumption is assimilated to the energy consumption of synaptic operations \cite{kundu2021spike}, thus other factors (such as neuron addressing in multiplexed architecture or memory accesses) are often neglected. Moreover, the models usually do not take into account some specific mechanisms, like membrane potential leakage, reset and biases integration. In \cite{SNN4FBnEB}, the authors proposed a metric based on synaptic operations only, and found great energy consumption savings for the SNN (up to $126\times$ more efficient than the FNN baseline). In \cite{lemaire2022synaptic} the authors demonstrated that such simplistic metrics were not always coherent with actual energy consumption of circuits on FPGA. When taking memory into account, another team \cite{rethinkingtheperformance} found equivalent energy consumptions for SNNs and FNNs using various topologies on CIFAR10. Additionally, reference \cite{davidson2021comparison} measured a theoretical maximum spike rate of 1.72 to guarantee energy savings in the SNN based on a detailed metric, accounting for synaptic operations, memory accesses and activation broadcast. Those energy consumption models are enlightening, but still fail to settle whether event-based processing is sufficient to increase energy efficiency. That is mostly because those metrics are too hardware specific, or do not take all significant sources of energy consumption into account.

In the present work, we propose a metric intended to be independent from low-level implementation choices, based on three main operations: neuron addressing, synaptic operations and memory accesses.

\section{Metrics}
\label{sec-METRICS}


\subsection{Operational cost}
\label{sec-METRICS-Ops}

In this section we define a metric to compute the number of ACC and MAC due to synaptic operations in SNNs and FNNs.

\subsubsection{Convolutional layers}

For a convolution layer, the number of filters is defined by $C_\text{out}$ and their size are noted $C_\text{in}\times H_\text{kernel}\times W_\text{kernel}$, where $C$, $H$ and $W$ stands for channel, height and width. The input and output of the layer are composed of a set of feature maps, with shapes $(C_\text{in}\times  H_\text{in}\times W_\text{in})$ and $(C_\text{out}\times  H_\text{out}\times W_\text{out})$ respectively. In the following we consider the padding mode ``\textit{same}'' and a stride $S$. The number of timesteps is noted $T$. The equations describing the number of MAC and ACC operations in FNNs and SNNs, for convolution layers, are summaried in Eq. \ref{eq-ACC_MAC_SNN_FNN}.

\begin{equation}
    \label{eq-ACC_MAC_SNN_FNN}
    \begin{split}
            \text{MAC}_\text{Conv-l}^\text{FNN} & = C_\text{out} \times H_\text{out} \times W_\text{out} \times C_\text{in} \times H_\text{kernel} \times W_\text{kernel}\\
            \text{ACC}_\text{Conv-l}^\text{FNN} & = C_\text{out} \times H_\text{out} \times W_\text{out} \\
            \text{MAC}_\text{Conv-l}^\text{SNN} & = T \times C_\text{out} \times H_\text{out} \times W_\text{out} \\
            \text{ACC}_\text{Conv-l}^\text{SNN} & = \theta_\text{l - 1} \times (\lceil \frac{H_\text{kernel}}{S} \rceil)\times (\lceil \frac{W_\text{kernel}}{S} \rceil) \times C_\text{out} \\ & + T \times C_\text{out} \times H_\text{out} \times W_\text{out} + \theta_\text{l} \\
    \end{split}
\end{equation}

In FNNs, the integration of dense input matrixes requires a MAC operation for each element of the convolution kernels. That is described in the first row of Eq. \ref{eq-ACC_MAC_SNN_FNN}. Additionally, the integration of synaptic biases as ACC operations, since it does not require multiplication with input activation. There is one bias per output neuron as shown in the second row of Eq. \ref{eq-ACC_MAC_SNN_FNN}.

On the other hand, for an SNN the input activations are sparse binary matrices. The number of operations of the layer $l$, depends on the number of input and output spikes of that layer, noted $\theta_\text{l - 1}$ and $\theta_\text{l}$. Since spikes are binary and, they are integrated via ACC operations in contrast with FNNs. Each input spike causes one ACC operation per element of each filter, as shown in the first term of the fourth row ($\text{ACC}_\text{Conv-l}^\text{SNN}$) of Eq. \ref{eq-ACC_MAC_SNN_FNN}. The second term accounts for the bias added to each membrane potential at each timestep. The third term accounts for the membrane potential reset whenever an output spike is generated. Additionally, SNNs may involve a membrane potential leakage (\textit{i.e.} LIF neurons), which is modeled by an additional MAC operations for each output neuron, which is repeated at each timestep. This is depicted in the last two rows of Eq. \ref{eq-ACC_MAC_SNN_FNN}.
\subsubsection{Fully-connected layers}
The same reasoning is applied to FC layers. For a given layer $l$ the number of input and output neurons is noted $N_{in}$ and $N_{out}$ respectively. The equations for the number of MAC and ACC operations attributable to synaptic operations in FC layers are summarized in Eq. \ref{eq-ACC_MAC_SNN_FNN_FC}.

\begin{equation}
    \label{eq-ACC_MAC_SNN_FNN_FC}
    \begin{split}
        \text{MAC}_\text{FC-l}^\text{FNN} & = N_\text{in} \times N_\text{out} \\
        \text{ACC}_\text{FC-l}^\text{FNN} & = N_\text{out} \\
        \text{MAC}_\text{FC-l}^\text{SNN} & = N_\text{out} \times T \\
        \text{ACC}_\text{FC-l}^\text{SNN} & = \theta_\text{l - 1} \times N_\text{in} \times N_\text{out} + T \times N_\text{out} \\
    \end{split}
\end{equation}

\subsection{Memory cost}
\label{sec-METRICS-Memory}

In order to provide an evaluation of the energy used by both the FNN and the SNN, multiple assumptions have to be made. Without these assumptions, results could vastly vary between different unconstrained hardware implementations. Each layer of the FNN is assumed to have its own local (non-shared) memory. As a result, activations need to be kept in memory (\textit{i.e.} I/O buffers) for all layers. The data flow between layers of the SNN is assumed to be sparse and asynchronous. Therefore, messages of incoming spikes must be buffered in a FIFO queue for each layer. Additionally, the SNN must keep the membrane potentials for all layers between timesteps. In both cases, we assume that all the memory is akin to local SRAM, including weights in order to support a reconfigurable architecture. Additionally, there is no local caching in a register bank. Only registers for the operands and a local accumulator are present and are excluded from this evaluation.
All data is assumed to be represented with the same number of bits, including the messages describing a spike.

\subsubsection{Memory accesses}
Data flowing from and to the memory is an important sink of energy.
We attempt to describe each read or write operation of both the FNN and the SNN for each layer in order to evaluate the possible energy savings from using an SNN, which is mainly a result of its sparsity.
Equations are provided for a single input for the FNN and a single timestep for the SNN.
\paragraph{Read operations to inputs}

For a formal Conv layer, each output position matches with read operations from all input channels and all positions for which the kernel applies. For a formal FC layer, the number of read operations for the input data is equal to the number of inputs $N_\text{in}$.
\begin{align}
    RdIn^\text{FNN}_\text{Conv} =& C_\text{in} \times C_\text{out} \times H_\text{out} \times W_\text{out}
    \times W_\text{kernel} \times H_\text{kernel}\\
    RdIn^\text{FNN}_\text{FC} =& N_\text{in}
\end{align}
For the SNN, the read operations in the queue directly depends on the number of incoming spikes $\theta_\text{l-1}$ and must be measured during inference:
\begin{equation}
    RdIn^\text{SNN} = \theta_\text{l-1}
\end{equation}
\paragraph{Read operations to parameters}

In a formal Conv layer, each output position matches with read operations for all the weights in all filters associated to all input channels. The biases generate additional reads for all output positions and all filters. For an FC layer, every weight corresponding to all output neurons $N_\text{out}$ and all inputs $N_\text{in}$. The biases cause additional read for all output neurons.
\begin{align}
    \begin{split}
    RdParam^\text{FNN}_\text{Conv} =&  (C_\text{in} \times W_\text{kernel} \times H_\text{kernel} + 1)\\
    \times& C_\text{out} \times W_\text{out} \times H_\text{out}\\
    RdParam^\text{FNN}_\text{FC} =& (N_\text{in} + 1) \times N_\text{out}
    \end{split}
\end{align}

In an spiking Conv layer, all received spikes $\theta_\text{l-1}$ will trigger a read for all output filters and all associated output positions (i.e. of the dimensions of the kernel). Biases for all output positions and filters must still be read. For an FC layer, the number of read operations for parameters is similar to an SNN except that weights are only read for all input spikes $\theta_\text{l-1}$. Biases must still be read for all output neurons $N_\text{out}$.
\begin{align}
    RdParam^\text{SNN}_\text{Conv} =& \theta_\text{l-1} \times C_\text{out} \times W_\text{kernel} \times H_\text{kernel}\nonumber\\
    +& C_\text{out} \times W_\text{out} \times H_\text{out} \\
    RdParam^\text{SNN}_\text{FC} =& \theta_\text{l-1} \times N_\text{out} + N_\text{out}
\end{align}

\paragraph{Read operations to potentials}

There is no membrane potential to update in an FNN so there is no associated read operation.

In a spiking Conv layer, the membrane potentials corresponding to all output positions affected by each input (i.e. of the dimensions of the kernel) in all filters must be read in order to update them. Biases need to be applied separately and therefore generate an additional read operation at each timestep for all output positions and all filters.
For FC layers, the potentials of all output neurons are read for each input. Biases are applied separately and therefore generate an additional read operation at each timestep for all output neurons.
\begin{align}
    RdPot_\text{Conv} =& \theta_\text{l-1} \times C_\text{out} \times W_\text{kernel} \times H_\text{kernel}\nonumber\\
    &+ C_\text{out} \times H_\text{out} \times W_\text{out}\\ 
    RdPot_\text{FC} =& (\theta_\text{l-1} + 1) \times N_\text{out} 
\end{align}

\paragraph{Write operations to outputs}

In an formal Conv layer, each output position in all filters require a write operation. For an FC layer, each output neuron require a write operation. In both cases, the output is assumed to be fully computed in the local accumulator, including bias, before being written to RAM.
\begin{align}
WrOut^\text{FNN}_\text{Conv} =& C_\text{out} \times H_\text{out} \times W_\text{out}\\
WrOut^\text{FNN}_\text{FC} =& N_\text{out}
\end{align}

For the SNN, the write operations to the queue directly depends on the number of generated spikes $N_\text{output}$ and must be measured during inference:
\begin{equation}
WrOut = N_\text{output} 
\end{equation}

\paragraph{Write operations to potentials}
There is no membrane potential to update in an FNN so there is no associated write operation.

In an spiking Conv layer, the membrane potentials corresponding to all output positions affected by each input (i.e. of the dimensions of the kernel) in all filters must be written to in order to update them. Additionally, the biases must also be written separately to the potentials for all output positions and all filters at each new timestep.
For an FC layer, the potentials of all output neurons must be written to for each input. Additionally, the biases must also be written to the potentials for all output neurons separately at each new timestep.
\begin{align}
    WrPot_\text{Conv} =& \theta_\text{l-1} \times C_\text{out} \times W_\text{kernel} \times H_\text{kernel}\nonumber\\
    &+ C_\text{out} \times H_\text{out} \times W_\text{out}\\ 
    WrPot_\text{FC} =& \theta_\text{l-1} \times N_\text{out} + N_\text{out} 
\end{align}

\subsection{Addressing in Sparse vs. Dense Convolutions}
\label{sec-METRICS-CONV}

In this subsection, we evaluate the cost of addressing in FNNs and SNNs. The first uses dense processing, in which all input synapses are stimulated at the same time. On the other hand, the second uses sparse processing, in which synapses are sparsely stimulated across time. Let us begin with convolution layers. In order to simplify the following matter, we consider convolutions with a ``\textit{same}'' padding (input and output feature maps of same sizes) and a stride of $S$. In FNNs, a kernel scans all its possible positions (depending on padding, stride...) on the input sample and generates a dense output feature-map. In such dense convolutions, computation is performed sequentially and addresses can be computed by incrementing only an index (by 1 or $S$) assuming the memory is contiguous and ordered the same way it is processed. Thus, one index runs through the input, one index runs through the output and one index runs through the weights. In SNNs, sparse convolutions are performed asynchronously upon reception of input spikes, thus the kernel positions (\textit{i.e.} output neuron addresses) must be calculated each time a spike is received. In a sparse representation, computation is performed non-sequentially with no prior knowledge of which output position is affected by an incoming spike. Computing the initial output position requires two multiplications. Thereafter, the computation of the remaining positions are computed by incrementing an index assuming the memory is contiguous and ordered as for FNNs. There is only one index running through the kernel weights. The cost of addressing in number ACC and MAC operations in spiking and formal convolution layer are summarized in Equation \ref{eq-ACC_MAC_Addr_CONV}.

\begin{equation}
    \label{eq-ACC_MAC_Addr_CONV}
    \begin{split}
            \text{ACC}_\text{Addr-Conv-l}^\text{FNN} & = C_\text{in}\times H_\text{in}\times W_\text{in} + C_\text{out}\times H_\text{out}\times W_\text{out} + C_\text{out} \\
            & \times H_\text{kernel}\times W_\text{kernel} \\
            \text{MAC}_\text{Addr-Conv-l}^\text{SNN} & = \theta_\text{l-1} \times 2\\
            \text{ACC}_\text{Addr-Conv-l}^\text{SNN} & = \theta_\text{l-1} \times C_\text{out}\times H_\text{kernel}\times W_\text{kernel}\\
    \end{split}
\end{equation}\\

Where $C$ are the numbers of channels, $W$ the widths and $H$ the heights, of respectively the inputs when the index is $\text{in}$, outputs when it is $\text{out}$ and kernels when it is $\text{kernel}$. $\theta_\text{l-1}$ is the number of input spikes.

The same reasoning is applied to fully-connected layers. In FNNs, one index runs through the input, and another index runs through both the output. In SNNs, one single index runs through the output upon receiving an input spikes. This yields Eq. \ref{eq-ACC_MAC_Addr_FC}.

\begin{equation}
    \label{eq-ACC_MAC_Addr_FC}
    \begin{split}
            \text{ACC}_\text{Addr-FC-l}^\text{FNN} & = N_\text{in} + N_\text{out}\\
            \text{ACC}_\text{Addr-FC-l}^\text{SNN} & = \theta_\text{l-1} \times N_\text{out}\\
    \end{split}
\end{equation}\\
Where $N_\text{in}$ and $N_\text{out}$ are resepctively the number of input and output neurons.

\subsection{Energy consumption metric}
\label{sec-METRICS-Energy}

In this section, we combine the equations obtained for computation, memory accesses and addressing in a global Energy evaluation metric. For this purpose, we multiply the energy cost of each operation by its number of occurrences, according to the metrics computed in subsections \ref{sec-METRICS-Ops}, \ref{sec-METRICS-Memory} and \ref{sec-METRICS-CONV}. Our model can be summarized as shown in Eq. \ref{eq-Energy_Metric1}:

\begin{equation}
    \label{eq-Energy_Metric1}
    E = E_\text{mem} + E_\text{ops+addr}
\end{equation}

Where $E^\text{mem}$ is the energy consumption of memory accesses, $E^\text{ops}$ that of synaptic operations, and $E^\text{addr}$ that of addressing mechanisms.

Those elements are computed based on the metrics proposed in the above subsections, as shown in Eq. \ref{eq-Energy_Metric_mem} for memory accesses, and Eq. \ref{eq-Energy_Metric_ops} for addressing and synaptic operations. Equations are not repeated for those two last elements since they are identical.

\begin{equation}
    \label{eq-Energy_Metric_mem}
    \begin{split}
        E_\text{mem}^\text{FNN} &= (RdIn^\text{FNN} + RdParam^\text{FNN}) \times E_\text{RdRAM}\\ &+ WrOut^\text{FNN} \times E_\text{WrRAM}\\
        E_\text{mem}^\text{SNN} &= (RdIn^\text{SNN} + RdParam^\text{SNN} + RdPot) \times E_\text{RdRAM} \\ &+ (WrOut^\text{SNN} + WrPot) \times E_\text{WrRAM}\\
    \end{split}
\end{equation}\\


With $E_\text{RdRAM}$ and $E_\text{WrRAM}$ the energy for a single read and a single write operation in RAM, respectively. In our computation, we assume that $E_\text{RdRAM}=E_\text{WrRAM}$ for simplicity purpose.

\begin{equation}
    \label{eq-Energy_Metric_ops}
    \begin{split}
        E_\text{ops+addr}^\text{FNN} &= (E_\text{ADD}+E_\text{MUL})\times  \text{MAC}_\text{ops+addr}^\text{FNN} \\
        &+ E_\text{ADD}\times  \text{ACC}_\text{ops+addr}^\text{FNN}\\
        E_\text{ops+addr}^\text{SNN} &= (E_\text{ADD}+E_\text{MUL})\times  \text{MAC}_\text{ops+addr}^\text{SNN} \\
        &+ E_\text{ADD}\times  \text{ACC}_\text{ops+addr}^\text{SNN}\\
    \end{split}
\end{equation}\\

Where $E_\text{ADD}$ and $E_\text{MUL}$ are the energy cost of single additions and multiplications respectively.

The energy consumption of single operations (addition, multiplication and memory accesses) are drawn from the literature \cite{jouppi2021ten} for 45nm CMOS technology. For addition and multiplication with 32-bit integers, we use respectively $0.1 pJ$ and $3.1 pJ$. For SRAM memory accesses, we compute a linear interpolation function based on 3 particular values : 8 kB ($10 pJ$), 32 kB ($20 pJ$) and 1 MB ($100 pJ$). This function enables to compute the energy cost of a memory access knowing the memory size (\textit{i.e.} knowing the network hyper-parameters).

\section{Methods}
    \subsection{Spike coding}\label{AA}
       The role of spike coding is to convert input pixels into spikes, which are in turn transmitted to an SNN for information processing. In the \emph{Direct encoding} scheme, a spiking neuron (e.g. LIF or IF) located in the first layer (encoding) of the network is fed with a constant or dynamic input 
       through $T$ timesteps. The first layer thus converts the analog input into spike trains. In this paper, we propose three different uses of such encoding which are detailed below.
        \subsubsection{Static Frame-based data encoding}
        The first method is Static Frame-based data encoding, in which the raw input data is directly broadcast to the first layer. The network input size is thus identical to that of the input sample, each of which is presented repeatedly during $T$ timesteps. This method is adapted to every type of conventional data. The major drawback of this method is that the whole processing must be repeated over several timesteps (bias integration, membrane leakage...).
         \subsubsection{Dynamic Frame-based data encoding} 
         The second method is Dynamic Frame-based data encoding, adapted to temporal signals. The input data is split into several chunks along the temporal dimension. The network input size is the same as the size of a chunk, which are presented successively at the input, one per timestep. This approach has two main benefits: reducing the input size and no longer considering timesteps mechanisms as an additional cost. 
        \subsubsection{Event-based data} 
        Event-based data does not require a specific encoding since events can be interpreted as spikes. However, in order to process event-based data in modern deep learning models, we need to convert them into a dense representation. To process events with FNNs, we simply sum all events occurring during the sample duration $d$ to reconstruct a single frame, containing integer values. 
        On the other hand, when using SNNs, we accumulate events over $T$ time windows (\textit{i.e.} timesteps) lasting $\Delta t = \frac{d}{T}$ to reconstruct frames. This type of representation is called a \textit{voxel grid} \cite{voxelgrid}, where each voxel represents a pixel and a time interval. We added the constraint that the accumulation is a simple OR operation: if at least one event is present during the time window then its value will be 1 \cite{loic_ijcnn2022}. This way, all of our event frames stays binary in order to leverage the efficiency of spiking neural networks running on specialized hardware \cite{abderrahmane2022spleat}. 
    \subsection{Organization of the output layer}\label{outputlayer}
    Both our FNN and SNN for static frame-based data use a traditional final fully-connected layer. On the other hand, our models for dynamic frame-based and event-based data use a specific final classification layer as the feature maps are not sufficiently reduced to be flattened before the final fully-connected layer.
    
    We followed the approach used in \cite{loic_ijcnn2022}. The output layer of our SNNs is simply composed of a batch normalization layer, a $1\times 1$ convolution outputting $num\_classes$ channels and a final layer of LIF neurons. The final predictions are then obtained by summing all output spikes first in the spatial dimension and time dimension. We therefore obtain a tensor with a spatial size of $1\times 1$ with $num\_classes$ channels, which is equivalent to the output of conventional fully-connected layers. The 1D convolution in the final layer enables to avoid the use of e.g. average pooling to reduce the spatial dimension as it would be incompatible with spikes computations. We used the same approach for the equivalent FNNs but without summation along the time axis since it does not exist.

    %
    

\section{Experiments and Results}

    
    \subsection{Datasets and Models}
    \label{sec-Results-CIFAR}
    
        \subsubsection{Static frame-based data}
            The CIFAR-10 dataset is made of 60000 32x32 RGB images representing 10 classes. For the SNN, CIFAR-10 samples are repeated as input over $T=4$ timesteps, following the static frame-based data encoding described in Section \ref{AA}. For this task, we use a VGG-16 architecture described in \cite{VGG}. We dropped the max pooling layers by using a stride of 2 in their preceding convolution and we added batch normalization layers after each convolution. 
    \subsubsection{Dynamic frame-based data}
     \label{sec-Results-GSC}


        
The Google Speech Commands V2 dataset is a dataset of audio signals sampled at 16~kHz composed of 1-second recordings of 35 spoken keywords. 
We performed data augmentation by randomly changing the speed of the raw audio signals.
The raw data are preprocessed to obtain images that can be fed to CNNs. We used 10 MFC Coefficients, FFT of size 1024, a window size of 640 with a hop of 320, and a padding of 320 on both sides. This results in a 48x10 image interpreted as 1D data truncated to 48 samples with 10 channels. For the SNN, we divided this temporal data in $T=2$ timesteps, each of size 24. To tackle this classification task, we designed a 4-layers CNN with the following topology: 48c3 - 48c3 - 96c3 - 35c1. Each convolutional layer has a stride of 1 and was followed by a batch normalization layer.

    \subsubsection{Event-based data}
        The Prophesee NCARS dataset \cite{ncars} is a classification dataset composed of 24k samples of length 100ms captured with a Prophesee GEN1 event camera mounted behind the windshield of a moving car. The samples represent either a car or background. 
        We resized all the samples to a size of $64 \times 64$ pixels using nearest neighbor interpolation to keep our inputs binary. For SNNs, we divided each sample in $T = 5$ timesteps, while all the events were summed into a single frame for the CNN. We proposed a variant of the classical Tiny VGG-11 architecture \cite{VGG} that uses 4 times fewer channels in each convolution layers, reducing the number of parameters and calculations. Once again, we dropped the max pooling layers by using a stride of 2 in their preceding convolution and we added batch normalization layers after each convolution. Finally, we replaced the final 3 fully-connected layers by our output layer described in Section \ref{outputlayer}.
        

\subsection{Results}
We trained our FNNs using PyTorch, and our SNNs using SpikingJelly \cite{SpikingJelly} with surrogate gradient learning. The models were trained over 50 epochs for GSC and NCARS, and 300 for CIFAR-10. All presented results represent the average over 5 runs. The performance of our networks were measured by their classification accuracy. We also measured the spike rate of our SNNs, corresponding to the average number of spikes per synapse in the network. Since computations are only performed when there is a spike, this has a direct impact on the SNN energy consumption within our metric.
These results are summarized in Table~\ref{tab:GSC_NCARS_CIFAR}. 

\begin{table}[ht]
\renewcommand{\arraystretch}{1.1}
\setlength\tabcolsep{3.0pt}
        \centering
        \caption{Accuracy and Activity comparisons between our proposed SNNs and CNNs on CIFAR-10, Google Speech Commands V2, and Prophesee NCARS.}
        \label{tab:GSC_NCARS_CIFAR}
\begin{tabular}{cccccc}
\hline
\textbf{Dataset}                   & \textbf{Network} & \textbf{\#Params}      & \textbf{Activation} & \textbf{Acc.} & \textbf{Spike Rate} \\ \hline
\multirow{2}{*}{\textbf{CIFAR-10}} & \multirow{2}{*}{VGG-16}                                          & \multirow{2}{*}{15.2M} & ReLU                                                            & 0.951         & --               \\
                                   &                                                                  &                        & IF                                                              & 0.884         & 0.10              \\ 
\multirow{2}{*}{\textbf{GSC}}      & \multirow{2}{*}{4-layers CNN}                                          & \multirow{2}{*}{29k}   & ReLU                                                            & 0.936         & --                \\
                                   &                                                                  &                        & LIF                                                             & 0.918         & 0.14              \\ 
\multirow{2}{*}{\textbf{NCARS}}    & \multirow{2}{*}{Tiny VGG-11}                                     & \multirow{2}{*}{356k}  & ReLU                                                            & 0.934         & --               \\
                                   &                                                                  &                        & LIF                                                             & 0.935         & 0.08              \\ \hline

\end{tabular}
\end{table}

For dynamic and event data, SNNs are able to reach equivalent or close accuracies to their FNN counterparts, a result never shown experimentally before on these datasets. On the CIFAR-10 dataset, our SNN reaches lower accuracy than the FNN. This is coherent with state of the art results, but should be improved in further works. Still, all of our SNNs reach these performance while having a very low spike rate, on average each neuron of a model spikes less than 0.14 times per inference for the three datasets.

\begin{figure*}[htb]
    \includegraphics[width=\textwidth]{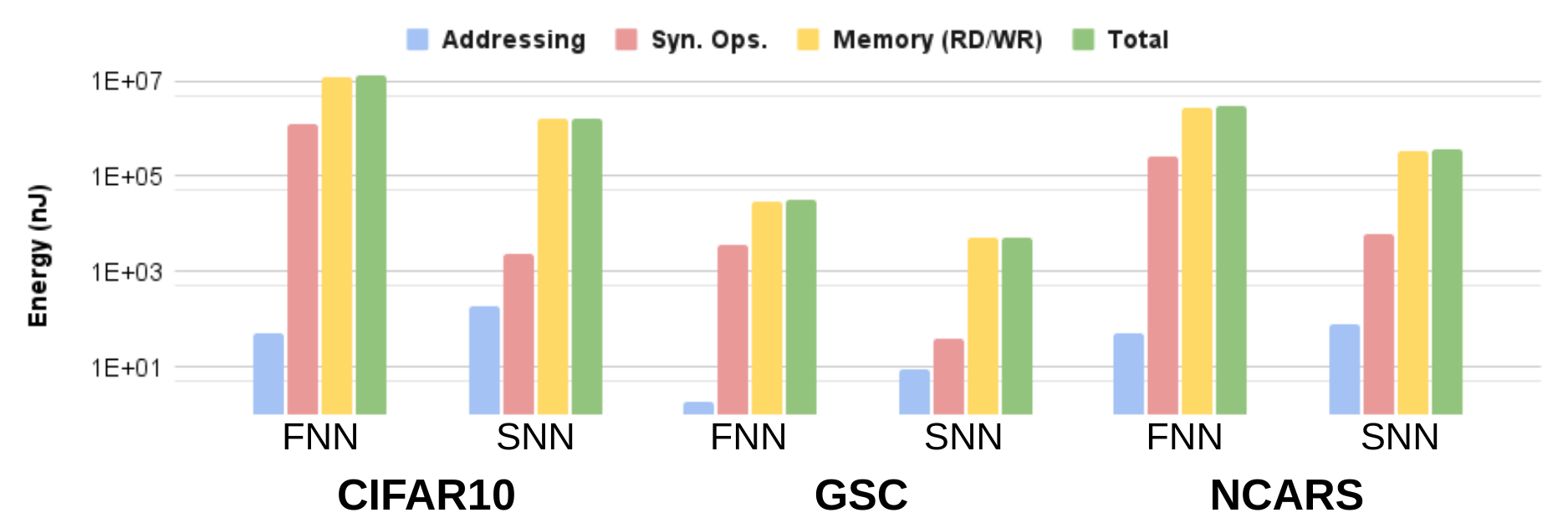}
    \caption{
    Estimation of energy consumption for 45nm CMOS technology.
    }
    \label{fig-energy_CIFAR_GSC_NCARS}
\end{figure*}

Using the metrics proposed in Section \ref{sec-METRICS-Memory}, we were able to precisely estimate the energy consumption of our models. In our model, I/O buffers between SNN layers are FIFOs able to store 1000 32-bit elements. This assumption has been validated through hardware simulation using SPLEAT architecture \cite{abderrahmane2022spleat}. On the other hand, the full feature maps are stored in SRAMs between FNN layers. 
The results are illustrated in Fig. \ref{fig-energy_CIFAR_GSC_NCARS} and detailed in Table \ref{tab:energy_all}.



\begin{table*}[htb]
\renewcommand{\arraystretch}{1.3}
\setlength\tabcolsep{5.0pt}
\centering
\caption{Energy consumption estimations of our SNNs and their equivalent FNNs on the 3 datasets for 45nm CMOS technology using the metrics proposed in Section \ref{sec-METRICS-Memory}.}        
\label{tab:energy_all}
\begin{tabular}{cccccccc}
\hline
\multirow{2}{*}{}                                                                          & \multirow{2}{*}{\textbf{}} & \multicolumn{2}{c}{\textbf{CIFAR10}} & \multicolumn{2}{c}{\textbf{GSC}}    & \multicolumn{2}{c}{\textbf{NCARS}}  \\
                                                                                           &                            & \textbf{SNN}      & \textbf{FNN}     & \textbf{SNN}     & \textbf{FNN}     & \textbf{SNN}     & \textbf{FNN}     \\ \hline
\multirow{5}{*}{\textbf{\begin{tabular}[c]{@{}c@{}}Memory\\ Accesses\\ (nJ)\end{tabular}}} & \textbf{Potentials}              & 1.12E+6           & --               & 3.50E+3          & --               & 2.69E+5          & --               \\
                                                                                           & \textbf{Weights}           & 4.60E+5           & 5.59E+6          & 1.73E+3          & 1.46E+4          & 8.10E+4          & 1.11E+6          \\
                                                                                           & \textbf{Bias}              & 2.79E+2           & 6.97E+1          & 6.00E+0          & 3.00E+0          & 3.71E+1          & 7.43E+0          \\
                                                                                           & \textbf{In/Out}            & 7.65E+2           & 6.09E+6          & 3.90E+1          & 1.50E+4          & 2.88E+2          & 1.54E+6          \\ \cline{2-8} 
                                                                                           & \textbf{Total}             & 1.58E+6           & 1.17E+7          & 5.27E+3          & 2.96E+4          & 3.50E+5          & 2.64E+6          \\ \hline
\multicolumn{2}{c}{\textbf{Synaptic Op. (nJ)}}                         & 2.41E+3           & 1.29E+06          & 4.08E+1          & 3.53E+3          & 6.15E+3          & 2.67E+5          \\ \hline
\multicolumn{2}{c}{\textbf{Addressing (nJ)}}                       & 1.97E+2           & 4.94E+1          & 9.28E+0          & 1.93E+0          & 5.05E+1          & 7.65E+1          \\ \hline
\multicolumn{2}{c}{\textbf{Total (nJ)}}                                                                                 & 1.58E+6  & 1.24E+7 &  5.32E+3 & 3.32E+4 & 3.57E+5 & 2.86E+6 \\ \hline

\multicolumn{2}{c}{\textbf{$E^\text{FNN}/E^\text{SNN}$}}                                & \multicolumn{2}{c}{\textbf{8.19}}    & \multicolumn{2}{c}{\textbf{6.22}}   & \multicolumn{2}{c}{\textbf{8.17}}   \\ \hline
\end{tabular}
\end{table*}

The total energy consumption is dominated by the cost of memory accesses, which is yet unduly neglected in most metrics of the literature. While SNNs have additional memory accesses for updating the neuron potentials, it also requires fewer memory accesses for the weights and the I/O. Moreover, the size of I/O buffers are often much smaller in SNNs than in FNNs, since the first only requires FIFOs of 1000 elements whereas the second requires storing the full feature maps.
The lower number of spikes combined with the absence of multiplication result in an energy consumption of synaptic operations two order of magnitude lower for SNNs than for FNNs. In the end, the total energy consumption of our SNNs is between 6.25 and 8.02 times lower than their FNN counterparts, a promising result for the implementation of SNNs on specialized hardware. 	
\section{Conclusion}

Neuromorphic Engineering is based on the assumption that event-based processing is the key to mimic the unparalleled energy efficiency of the biological brain. However, this assumption remains to be proven. 
The goal of this work was to settle this question through a generic and accurate energy estimation metric, independent from low-level implementation choices and hardware targets. Our proposed analytical model is based on three types of operations occurring in hardware neural network implementations: synaptic operations, memory accesses and addressing mechanisms. This metric was applied to three datasets representative of three characteristic data types. 
In all three cases, spiking implementation could bring major energy savings compared to 
formal ones as our SNNs are respectively 8.19x, 6.22x and 8.17x more efficient on CIFAR-10, GSC and NCARS, while producing near state-of-the-art accuracy for the last two. 

Future works will include a confrontation of those results with actual energy consumption measurements, using our own SNN hardware architecture (SPLEAT \cite{abderrahmane2022spleat}) and other state-of-the-art deep learning accelerators. Moreover, further work is required on our CIFAR-10 model, in order to reach state-of-the-art accuracy and thus increase the fairness of the comparison. Additionally, we will study the impact of various quantization schemes on energy consumption.






\bibliography{bibliography}

\begin{thebibliography}{10}

\bibitem{confortingml}
L.~Khacef, N.~Abderrahmane, and B.~Miramond, ``Confronting machine-learning
  with neuroscience for neuromorphic architectures design,'' in {\em Int. Joint
  Conf. on Neural Netw.}, 2018.

\bibitem{NxTF}
B.~Rueckauer {\em et~al.}, ``Nxtf: An api and compiler for deep spiking neural
  networks on intel loihi,'' 2021.

\bibitem{lemaire2022modelisation}
E.~Lemaire, {\em Mod{\'e}lisation et exploration d'architectures
  neuromorphiques pour les syst{\`e}mes embarqu{\'e}s haute-performance}.
\newblock PhD thesis, Univ. C{\^o}te d'Azur, 2022.

\bibitem{lemaire2022synaptic}
E.~Lemaire, B.~Miramond, S.~Bilavarn, H.~Saoud, and N.~Abderrahmane, ``Synaptic
  activity and hardware footprint of spiking neural networks in digital
  neuromorphic systems,'' {\em ACM Trans. on Embedded Computing Syst.}, 2022.

\bibitem{kheradpisheh2020temporal}
S.~R. Kheradpisheh and T.~Masquelier, ``Temporal backpropagation for spiking
  neural networks with one spike per neuron,'' {\em Int. J. of Neural Syst.},
  vol.~30, no.~06, p.~2050027, 2020.

\bibitem{zimmer2019technical}
R.~Zimmer, T.~Pellegrini, S.~F. Singh, and T.~Masquelier, ``Technical report:
  supervised training of convolutional spiking neural networks with pytorch.''
  2019.

\bibitem{pellegrini2021low}
T.~Pellegrini, R.~Zimmer, and T.~Masquelier, ``Low-activity supervised
  convolutional spiking neural networks applied to speech commands
  recognition,'' in {\em 2021 IEEE Spoken Language Technology Workshop (SLT)},
  pp.~97--103, IEEE, 2021.

\bibitem{amir2017low}
A.~Amir {\em et~al.}, ``A low power, fully event-based gesture recognition
  system,'' in {\em IEEE Conf. on Comput. Vision and Pattern Recognition},
  pp.~7243--7252, 2017.

\bibitem{ijcai2021-321}
J.~Ding, Z.~Yu, Y.~Tian, and T.~Huang, ``Optimal {ANN-SNN} conversion for fast
  and accurate inference in deep spiking neural networks,'' in {\em Int. Joint
  Conf. on Artif. Intell.}, pp.~2328--2336, 2021.

\bibitem{neftci}
E.~Neftci, H.~Mostafa, and F.~Zenke, ``Surrogate gradient learning in spiking
  neural networks: Bringing the power of gradient-based optimization to spiking
  neural networks,'' {\em IEEE Signal Process. Magazine}, vol.~36, pp.~51--63,
  2019.

\bibitem{SNN4FBnEB}
S.~Barchid, J.~Mennesson, J.~Eshraghian, C.~Djéraba, and M.~Bennamoun,
  ``Spiking neural networks for frame-based and event-based single object
  localization,'' 2022.

\bibitem{kundu2021spike}
S.~Kundu, G.~Datta, M.~Pedram, and P.~A. Beerel, ``Spike-thrift: Towards
  energy-efficient deep spiking neural networks by limiting spiking activity
  via attention-guided compression,'' in {\em Proceedings of the IEEE/CVF
  Winter Conference on Applications of Computer Vision}, pp.~3953--3962, 2021.

\bibitem{rethinkingtheperformance}
L.~Deng {\em et~al.}, ``Rethinking the performance comparison between {SNNs}
  and {ANNs},'' {\em Neural Netw.}, vol.~121, pp.~294--307, 2020.

\bibitem{davidson2021comparison}
S.~Davidson and S.~B. Furber, ``Comparison of artificial and spiking neural
  networks on digital hardware,'' {\em Frontiers in Neuroscience}, vol.~15,
  p.~651141, 2021.

\bibitem{jouppi2021ten}
N.~P. Jouppi {\em et~al.}, ``Ten lessons from three generations shaped
  google’s tpuv4i: Industrial product,'' in {\em ACM/IEEE Annu. Int. Symp. on
  Comput. Architecture}, pp.~1--14, 2021.

\bibitem{voxelgrid}
P.~Bardow, A.~J. Davison, and S.~Leutenegger, ``Simultaneous optical flow and
  intensity estimation from an event camera,'' in {\em IEEE Conf. on Comput.
  Vision and Pattern Recognition}, pp.~884--892, 2016.

\bibitem{loic_ijcnn2022}
L.~Cordone, B.~Miramond, and P.~Thierion, ``Object detection with spiking
  neural networks on automotive event data,'' in {\em Int. Joint Conf. on
  Neural Networks}, 2022.

\bibitem{abderrahmane2022spleat}
N.~Abderrahmane, B.~Miramond, E.~Kervennic, and A.~Girard, ``Spleat: Spiking
  low-power event-based architecture for in-orbit processing of satellite
  imagery,'' in {\em Int. Joint Conf. on Neural Networks}, 2022.

\bibitem{VGG}
K.~Simonyan and A.~Zisserman, ``Very deep convolutional networks for
  large-scale image recognition,'' in {\em Int. Conf. on Learning
  Representations}, 2015.

\bibitem{ncars}
A.~Sironi, M.~Brambilla, N.~Bourdis, X.~Lagorce, and R.~Benosman, ``Hats:
  Histograms of averaged time surfaces for robust event-based object
  classification,'' in {\em IEEE Conf. on Comput. Vision and Pattern
  Recognition}, June 2018.

\bibitem{SpikingJelly}
W.~Fang, Y.~Chen, J.~Ding, D.~Chen, Z.~Yu, H.~Zhou, Y.~Tian, and other
  contributors, ``Spikingjelly.''
  \url{https://github.com/fangwei123456/spikingjelly}, 2020.
\newblock Accessed: 2022-07-29.

\end{thebibliography}
\bibliographystyle{ieeetr}

\end{document}